\documentclass[aps, floats,twocolumn,superscriptaddress,amsmath,amssymb]{revtex4}
\usepackage[varg]{txfonts}
\usepackage[final]{graphicx}
\DeclareGraphicsExtensions{.eps,.ps,.pdf}
\usepackage{amsmath,amssymb,bbold,bm}
\newcommand\vex[1]{\mathbf{#1}}

\newcommand\bvarphi{\bm{\varphi}}
\def\d{\mathrm{d}}
\def\id{\mathbb{1}}

\def\sslash#1{\setbox0=\hbox{$#1$}			
   \dimen0=\wd0                                		
   \setbox1=\hbox{/} \dimen1=\wd1  	 		
   \ifdim\dimen0>\dimen1                               	
      \rlap{\hbox to \dimen0{\hfil / \hfil}} 	  	
      #1                                      			
   \else                                        			
      \rlap{\hbox to \dimen1{\hfil$#1$\hfil}}   		
      \hbox{/} 	                              			
   \fi}   
\def\slash#1{\hbox{$#1$\kern-0.35em\raise0.1ex\hbox{/}}}

\begin{document}

\title{Fractional statistics of topological defects in graphene and related structures}
\author{B. Seradjeh}
\affiliation{Department of Physics and Astronomy, University of British Columbia,
Vancouver, BC, Canada V6T 1Z1}
\author{M. Franz}
\affiliation{Department of Physics and Astronomy, University of British Columbia,
Vancouver, BC, Canada V6T 1Z1}
\affiliation{Kavli Institute for Theoretical Physics, University of California, Santa Barbara, CA 93106}
\date{\today}

\begin{abstract}
We show that fractional charges bound to topological defects in the recently proposed 
time-reversal-invariant models on honeycomb and square lattices obey fractional 
statistics. The effective low-energy description is given in terms of a `doubled' 
level-2 Chern-Simons field theory, which is parity and time-reversal invariant 
and implies two species of semions (particles with statistical angle 
$\pm\pi/2$) labeled by a new emergent quantum number that we identify as the fermion axial charge.

\end{abstract}

\maketitle

\emph{Introduction} -- When the excitations of a many-body system carry electric
charge that is smaller than the charge of its constituent particles
(e.g., electrons) the charge is said to be {\em fractionalized}.  This 
phenomenon is known to occur in the fractional quantum 
Hall (FQH) liquids~\cite{Lau83b}, quintessential strongly correlated systems
with broken time-reversal symmetry, where fractionally-charged excitations also 
obey fractional exchange statistics~\cite {AroSchWil84a}. Recently, two model 
systems have been introduced on the honeycomb and square lattices~\cite{HouChaMud07a,SerWeeFra07a}  that exhibit charge fractionalization 
{\em without breaking of the time-reversal symmetry}. These models, in essence, 
generalize the concept of fractionalization in polyacetylene~\cite{SuSchHee79a} 
to two dimensions and, remarkably, can be considered weakly correlated. 
The experience with FQH systems suggests that 
the exchange statistics of these fractionally charged excitations 
could be anomalous. This question is interesting for several 
reasons. First, a very general argument can be made~\cite{kivelson1} that would 
seem to prohibit the existence of anyons in  systems that obey time-reversal 
symmetry. Second, fractional statistics have recently 
captured attention due to their relevance to topologically protected quantum 
information processing 
\cite{kitaev1}. Since the honeycomb lattice is found in natural graphene 
\cite{geim1}, and the square lattice model could be realized in artificially
engineered structures~\cite{weeks1}, the possibility of realizing anyons in 
time-reversal invariant systems has both theoretical and practical 
significance.

In this Letter we construct the low-energy effective theory for the fractional 
particles
in models~\cite{HouChaMud07a,SerWeeFra07a}. We find that they are indeed anyons, 
albeit of a very special kind, described by a {\em doubled} 
U(1)$_2\times\overline{\rm U(1)}_2$ Chern-Simons (CS) 
theory previously discussed by Freedman {\em et al.}~\cite{freedman1}. In its 
topological sector the theory contains two species of semions, which transform 
into each other under parity and time reversal, thus escaping the constraints 
imposed by the argument of Ref.~\onlinecite{kivelson1}. Systems under 
consideration here~\cite{HouChaMud07a,SerWeeFra07a} represent the first 
explicit example of models for which such a gauge structure emerges as the
low-energy effective theory.

\emph{Fractional charge} -- 
The low-energy theory for fermions on the graphene honeycomb 
lattice~\cite{semenoff} and the square lattice threaded with $\pi$ flux per 
plaquette~\cite{hof1} is the Dirac Lagrangian
\begin{equation}\label{eq:l1}
\mathcal{L} = \bar\psi\left(i\slash\partial+m e^{-i\chi\gamma_5}\right)\psi
\end{equation}
where $\slash z=\gamma_\mu z_\mu$, $\mu=0,1,2$, $\gamma_\mu$ are $4\times 4$ 
Dirac matrices in the Weyl representation and $\psi$ is a four-component Dirac 
spinor whose components index two Dirac points (`valleys') and two sublattices. 
The mass $m$ arises from the dimerization of hopping amplitudes introduced in 
Refs.~\onlinecite{HouChaMud07a,SerWeeFra07a}. The phase $\chi$ describes the
direction of the dimerization pattern. Hou {\em et al.}~\cite{HouChaMud07a} 
made a  remarkable discovery that a topological defect (a vortex 
in $\chi$) binds fractional charge $\pm e/2$.
Determination of this fractional charge relies on establishing the existence of
unpaired zero modes in the solution of the associated Dirac equation~\cite{JacReb76a,JacRos81a,Wei81a}. For our discussion it is useful to deduce the fractional charge by a method that
does not rely on zero modes but instead exploits the long-distance behavior
of Dirac fermions in topologically non-trivial backgrounds. 

\begin{figure}
\includegraphics[width=8cm]{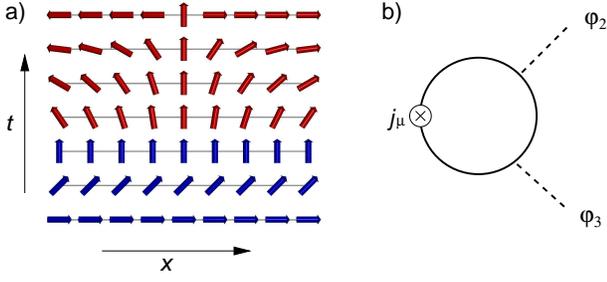}
\caption{(Color online) (a) Adiabatic formation of a vortex from a uniform,
vortex-free configuration. The arrows represent the direction of $\hat\bvarphi$
along the $y=0$ cut through the vortex center. (b) Feynman diagram used to
calculate the fermion current. Solid lines represent the fermion propagator
$G(p)=(\slash{p}-m)^{-1}$.
}
\label{fig1}
\end{figure}

The idea here, originally due to Goldstone and Wilczek~\cite{GolWil81a}, is to 
follow the flow of charge as we introduce a vortex into $\chi$ by adiabatically 
deforming the order parameter, starting from a uniform 
configuration, e.g., $\chi(x)=0$. This adiabatic insertion is made possible by
temporarily enlarging the parameter space of masses by adding to the Lagrangian 
(\ref{eq:l1}) a ``$\gamma_3$-mass'' term, $m_3\bar\psi\gamma_3\psi$. Physically this
term corresponds to staggered on-site potential $\pm m_3$ for the fermions 
on two sublattices of the square or honeycomb lattice. Such a term could in
principle appear in the physical Lagrangian but we add it here by hand to 
enlarge the symmetry of the order parameter from U(1) to O(3). Let us denote 
this O(3) order parameter by the vector 
\begin{equation}\label{eq:pol}
\bvarphi(x)=(m\cos\chi,m\sin\chi,m_3)
\end{equation}
 with a fixed length 
$\bvarphi^2\equiv M^2=m^2+m_3^2$ and in the direction of the unit vector
$\hat\bvarphi(x)$. Now we can adiabatically create a vortex in 
$\hat\varphi_1+i\hat\varphi_2$ by first rotating $\hat\bvarphi$ from the initial
uniform state ``up'' or ``down'' to  $(0,0,\mathrm{sgn}(m_3))$ and then 
flattening to the 
$m_3=0$ plane away from the center as illustrated 
in Fig.~\ref{fig1}(a). This creates a \emph{meron} (half a skyrmion) in 
$\hat\bvarphi$ with the core pointing in the $(0,0,\mathrm{sgn}(m_3))$ 
direction. On the lattice we can always take the limit of zero core size, thus
recovering an ordinary U(1) vortex. In the continuum description under 
consideration here, however, there always remains a single point (coincident
with the singularity in $\chi$) where $m_3$ retains a nonzero value, $\pm M$.
This value distinguishes between the two different ways of creating a vortex and will be seen below to have physical implications.

The utility of this procedure lies in the fact that the formation of a vortex is
completely {\em smooth} and we can thus calculate the charge accumulated 
during this adiabatic process by a perturbative loop-expansion of the 
fermion current operator. 
On symmetry and dimensional grounds we may expect that the fermion current 
will take the form of the conserved topological current in the O(3) nonlinear 
$\sigma$ model, 
\begin{equation}\label{eq:O3cur}
j_\mu^{\mathrm{top}}=\frac1{8\pi}\epsilon_{\mu\nu\lambda}\epsilon_{abc}\hat\varphi_a\partial_\nu\hat\varphi_b\partial_\lambda\hat\varphi_c,
\end{equation}
up to an overall prefactor. As in Ref.~\onlinecite{GolWil81a} the correctness of
this anticipation and the prefactor can be established by computing the 
expectation value of the fermion current $j_\mu=\bar\psi\gamma_\mu\psi$ in 
response to a small perturbation $\bvarphi(x)$ imposed on top of a uniform 
background $\bvarphi_0$, which we take to lie in the $(1,0,0)$ 
direction. A straightforward evaluation of the diagram in Fig.~\ref{fig1}(b)
yields $\langle j_\mu\rangle=(M/8\pi|M|^3)\epsilon_{\mu\nu\lambda}\partial_\nu\varphi_2\partial_\lambda\varphi_3$. Considering other permutations of external fields
in the above diagram we identify
$\langle j_\mu\rangle=j_\mu^{\mathrm{top}}$ for the general background field.

To calculate the charge associated with the meron configuration let us substitute 
$\bvarphi(x)$ from Eq.~(\ref{eq:pol}) into (\ref{eq:O3cur}) to obtain
\begin{equation}\label{eq:curpol}
\langle j_\mu\rangle =-{1\over 4\pi |M|}\epsilon_{\mu\nu\lambda}(\partial_\nu m_3)(\partial_\lambda\chi).
\end{equation}
The total charge 
$Q=\int\d^2\vex{x} \langle j_0\rangle = (4\pi|M|)^{-1}\int\d^2\vex{x}
m_3(\vex{x})\epsilon_{ij}\partial_i\partial_j\chi$ where we have used integration
by parts and the fact that  $m_3(\vex{x})=0$ far from the meron center. 
Recalling that for a static meron centered at the origin 
$\epsilon_{ij}\partial_i\partial_j\chi(\vex{x})=
2\pi n\delta(\vex{x})$ with $n$ the integer vorticity of $\chi$
we obtain $Q={1\over 2}n\mathrm{sgn}(m_3(\vex 0))$. This result has the satisfying
property of not depending on the exact spatial profile of $m_3$ but only
on its asymptotic value at the vortex center. This is in complete agreement with
Refs.\ ~\cite{JacReb76a,JacRos81a,HouChaMud07a,JacPi07a} when we recall that
the charge of a $n=\pm 1$ vortex can be either $+{1\over 2}$ or $-{1\over 2}$
depending on whether or not is the zero mode occupied by an electron. In our
present construction the sign of $m_3$ is seen to play the role of the electron
occupancy, despite the fact that the zero mode never enters our discussion.

For future reference we also record the form of $\langle j_\mu\rangle$ in the 
limit of infinitely small meron core, i.e., $m_3(\vex{x})=0$ everywhere except 
$m_3(\vex 0)=\pm M$. Eq.~(\ref{eq:curpol}) becomes
\begin{equation}\label{eq:curpol2}
\langle j_\mu\rangle ={1\over 4\pi}\mathrm{sgn}(m_3)\epsilon_{\mu\nu\lambda}
\partial_\nu\partial_\lambda\chi.
\end{equation}
This generalizes naturally to the case of multiple vortices with $m_3=\pm M$ at
each vortex center.

\emph{Effective theory} -- We now turn to the discussion of statistics. Our 
strategy will be to find the  low-energy effective theory for vortices in the 
Lagrangian (\ref{eq:l1}).
For the subsequent considerations it will be useful to modify the Lagrangian 
by including a coupling of the fermions to two gauge fields, 
\begin{equation}\label{eq:l2}
\mathcal{L} = \bar\psi \left(i\slash\partial - \slash A + \gamma_5 \slash B + 
m e^{-i\chi\gamma_5}\right)\psi.
\end{equation}
$A_\mu$ and $B_\mu$ couple minimally to the electric and the axial fermion 
currents, 
$j_\mu=\bar\psi\gamma_\mu\psi$ and $j_{5\mu}=\bar\psi\gamma_\mu\gamma_5\psi$, 
respectively, and can be therefore identified as the ordinary electromagnetic 
field and
the chiral gauge field introduced in Ref.~\onlinecite{JacPi07a}.
In what follows we treat $A$ and $B$ as static external fields that help us
keep track of the charge content of various fields.

We note that the coupling between vortices and fermions can be written as 
$m(\bar\psi_+e^{i\chi}\psi_-+\bar\psi_-e^{-i\chi}\psi_+)$ where 
$\psi_\pm=\frac12(1\pm \gamma_5)\psi$ are the chiral components of the Dirac 
fermion. This suggests that in order to focus on the vortex degrees of freedom
we perform a singular gauge transformation
 $\psi_\pm\to e^{\pm i\chi_\pm}\psi_\pm$, where we have defined an (as yet) 
arbitrary partitioning $\chi=\chi_++\chi_-$ of the vortex phase, such that upon 
encircling a vortex the spinor field remains single-valued~\cite{FraTes01a}.
(This simply means that the phase field associated with any given vortex is 
assigned to either $\chi_+$ or $\chi_-$.) 
Under this transformation the Lagrangian becomes
\begin{equation}\label{eq:l3}
\mathcal{L} = \bar\psi \left(i\slash\partial - \slash a + \gamma_5 \slash b + 
m \right)\psi,
\end{equation}
with shifted gauge fields
\begin{subequations}\label{eq:shift}
\begin{eqnarray}
a_\mu &=& A_\mu+\frac12\partial_\mu(\chi_+-\chi_-), \label{eq:shift1}\\
b_\mu &=& B_\mu+\frac12\partial_\mu(\chi_++\chi_-). \label{eq:shift2}
\end{eqnarray}
\end{subequations}
A transformation of this type has been previously employed in the studies of the 
vortex state of a $d$-wave superconductor~\cite{FraTes01a}. Its chief advantage 
is that it recasts a somewhat unwieldy Lagrangian (\ref{eq:l1}) with a twisted
mass term in the form of a gauge theory (\ref{eq:l3}) which can be analyzed by
standard field-theory methods. Specifically, our goal is to integrate
out the fermi fields in (\ref{eq:l3}) to arrive at the effective action
in terms of the gauge fields which now encode the vortex degrees of freedom. 
As is usually the case this procedure encounters 
ultraviolet divergences. These must be regularized in a manner that is 
consistent with the symmetries of the underlying lattice model. In the following
analysis we shall pay particular attention to the time reversal and parity which 
we expect to be preserved in the low-energy theory.

\emph{Symmetries} -- Besides Lorenz and gauge invariance, the low-energy 
theory also respects discrete symmetries. At $m=0$ 
parity, $(t, x, y) \stackrel{\mathcal{P}}{\to} (t, -x, y)$, acts as 
$\psi \stackrel{\mathcal{P}}{\to} P \psi$ with $P\in\gamma_1\mathbf{u}(2)$
 where $\mathbf{u}(2)$ is the group generated by 
$\{\id, i\gamma_3, \gamma_5, \gamma_3\gamma_5\}$.  Charge conjugation, 
$\mathcal{C}$, can be similarly worked out to be $\psi\stackrel{\mathcal{C}}{\to}C\psi^*$ with $C\in\gamma_2\mathbf{u}(2)$. The antiunitary time-reversal 
operation, $(t, x, y)\stackrel{\mathcal{T}}{\to} (-t, x, y)$, is given 
by $\psi\stackrel{\mathcal{T}}{\to}T\psi$ with $T\in\gamma_1\mathbf{u}(2)$. 
The Lagrangian is odd under the unitary operation 
$(t, x, y)\stackrel{\mathcal{S}}{\to} (-t, x, y)$ that sends 
$\psi\stackrel{\mathcal{S}}{\to}S\psi$ with $S\in\gamma_0\mathbf{u}(2)$. That 
is, $\mathcal{S}$ anticommutes with the Hamiltonian. This is identified on the 
lattice as a sublattice symmetry~\cite{HouChaMud07a} which renders the energy 
spectrum symmetric around zero.

The sense of the vortex is switched under parity: 
$\chi\stackrel{\mathcal{P}}{\to}-\chi$. This means that 
$(b_0,b_1,b_2)\stackrel{\mathcal{P}}{\to}(-b_0,b_1,-b_2)$. The gauge field $a$ 
must, however, behave as a vector like the usual electromagnetic field, 
$(a_0,a_1,a_2)\stackrel{\mathcal{P}}{\to}(a_0,-a_1,a_2)$. So, we require that 
$\chi_\pm\stackrel{\mathcal{P}}{\to}-\chi_\mp$. Similarly we find 
$\chi_\pm\stackrel{\mathcal{C}}{\to}-\chi_\pm$ and 
$\chi_\pm\stackrel{\mathcal{T}}{\to}+\chi_\mp$. The latter implies 
$(b_0,b_1,b_2)\stackrel{\mathcal{T}}{\to}(-b_0,b_1,b_2)$ and 
$(a_0,a_1,a_2)\stackrel{\mathcal{T}}{\to}(a_0,-a_1,-a_2)$.
These conditions also uniquely determine the operation of $\mathcal{P}, 
\mathcal{C}$, $\mathcal{T}$ and $\mathcal{S}$ on the fermi fields in the 
presence of vortices ($m\neq0$) to be
\begin{equation}
P = \gamma_1\gamma_5;~~ C = \gamma_2\gamma_5;~~ T=\gamma_1\gamma_5;~~S=\gamma_0\gamma_3.
\end{equation}
It follows that the bilinear $\bar\psi\gamma_3\psi$ is even under any 
operation, while $\bar\psi\gamma_3\gamma_5\psi$ is even under $\mathcal{S}$ and 
odd under $\mathcal{P}$, $\mathcal{C}$ and $\mathcal{T}$. (In fact, the latter 
is true in the absence of vortices as well.)

\emph{Topological terms} -- Armed with the above analysis
we can now ask what topological terms are allowed by symmetry in the 
low-energy effective action for $a$ and $b$. It is easy to see that 
conventional CS terms 
$a\cdot(\partial\times a)$ and $b\cdot(\partial\times b)$ break both 
$\mathcal{P}$ and $\mathcal{T}$
 and are thus prohibited. One can, however, construct a
{\em mixed} CS term,
\begin{equation}\label{eq:lcs}
\mathcal{L}_{\rm CS} = {\kappa\over 4\pi}a\cdot(\partial\times b)
\end{equation}
which obeys $\mathcal{P}$ and $\mathcal{T}$ and is thus allowed. The value of 
the coefficient $\kappa$ is tied to the value of fractional charge. The simplest
 way to see this is to note that in view of Eq.~(\ref{eq:shift1}) varying the 
action $\mathcal{S}_{\rm CS}=
\int \d t \d^2\vex x \mathcal{L}_{\rm CS}$ with respect to $A_\mu$ gives the fermion
current, 
\begin{equation}\label{eq:curpol3}
\langle j_\mu\rangle ={\delta \mathcal{S}_{\rm CS}\over \delta A_\mu}\biggr|_{A,B=0}
={\kappa\over 8\pi}\epsilon_{\mu\nu\lambda}
\partial_\nu\partial_\lambda\chi.
\end{equation}
Consistency with Eq.~(\ref{eq:curpol2}) then requires $\kappa=2\mathrm{sgn}(m_3)$.

The mixed CS term (\ref{eq:lcs}) can also be obtained more directly from the
Lagrangian (\ref{eq:l3}) if we regularize it by adding a Pauli-Villars 
mass term. Based on our discussion of symmetries we choose to add the $\gamma_3$-mass term, which preserves both $\mathcal{T}$ and
 $\mathcal{P}$ but breaks $\mathcal{S}$. The physical amplitudes are found in a 
standard perturbative expansion using the Pauli-Villars subtraction, 
$\mathcal{A}_{\mathrm{phys}}=\mathcal{A}_{\mathrm{reg}}(m_3=0)-\mathcal{A}_{\mathrm{reg}}(m_3\to\infty)$.
Integrating out the Dirac fermions to one-loop order is a lengthy but ultimately
straightforward exercise that yields~\cite{ser1}
\[
\mathcal{L}_{\mathrm{eff}} = -\frac{\pi}{12|m|}(\partial\times a)^2+
\frac{|m|}{2\pi}b^2+\frac{\mathrm{sgn}(m_3)}{2\pi}a\cdot(\partial\times b).
\]
The last term is the mixed CS term obtained before. The $(\partial\times a)^2$
term
describes the expected dielectric response of the Dirac medium. In the absence
of the chiral gauge field the $b^2$ term becomes simply $(\partial\chi)^2$.
This reflects the cost of spatial and temporal variations in $\chi(x)$ and 
encodes the usual logarithmic interaction between vortices. If the chiral gauge
field is present and described at the bare level by a 
Maxwell term, then the interaction between vortices is exponentially screened
at long distances, as in ordinary type-II superconductors. 
Note that $m$ appears as the charge of $a$ and at the same time as the mass of 
$b$. Therefore, by tuning $m\to0$ we could expect to find a phase where $a$ is 
massive and $b$ is soft, hence the role of axial and regular currents is 
reversed. This would be a superconducting phase.

\emph{Exchange statistics} -- We now analyze the implications of the mixed CS 
term (\ref{eq:lcs}) for the vortex statistics. To this end we substitute $a$ and 
$b$ from Eqs.\ (\ref{eq:shift}) into $\mathcal{L}_{\rm CS}$ and set $A=B=0$,
\begin{equation}\label{eq:lcs2}
\mathcal{L}_{\mathrm{CS}}=\frac{\kappa}{16\pi}(u_+\cdot\partial\times u_+
-u_-\cdot\partial\times u_-),
\end{equation}
where we defined $u_{\pm\mu}=\partial_\mu\chi_\pm$. 
It is now easy to understand the exchange statistics. Take two vortices 
at $\mathbf{x}_1$ and $\mathbf{x}_2$ and let the second one go on a path 
$C_2$ around the first one, which remains static. Let us assume they both 
belong to the same \mbox{$\pm$-partition} and $\mathrm{sgn}(m_3)>0$. In Eq.~(\ref{eq:lcs2}) we may then write $u_\pm=u_1+u_2$ and $u_\mp=0$ where $\frac{1}{2\pi}(\partial\times u_k)_\mu = (\d x_\mu/\d t)\delta^{(2)}(\mathbf{x}-\mathbf{x}_k(t))$ is the current density of the $k$-th vortex. The topological phase $e^{2i\theta}$ that is 
accumulated in this process can be calculated from the 
$u_1\cdot\partial\times u_2$ cross term in Eq.~(\ref{eq:lcs2}):
\begin{eqnarray}
2\theta &=& \pm\frac12\int\d^3 x \delta^{(2)}(\mathbf{x}-\mathbf{x}_2(t))\frac{\d x_\mu}{\d t}u_{1\mu} \nonumber \\
&=& \pm\frac12 \oint_{C_2} \d\mathbf{x}_2\cdot\mathbf{u}_1 = \pm\pi.
\label{theta}
\end{eqnarray}
This means that two such vortices behave as {\em semions} with the statistical 
angle $\theta=\pm\pi/2$. Alternatively, if we have two vortices in two 
different $\pm$-partitions there is no cross term and they will be 
mutual bosons.

How can the assignment of vortices into the seemingly arbitrary 
\mbox{$\pm$-partitions} produce observable effects? The answer lies in the 
realization that this assignment actually entails a genuine physical 
distinction between vortices in the two partitions. This can be seen by 
computing the axial current by varying $\mathcal{S}_{\rm CS}$ with respect to 
$B_\mu$, as in Eq.~(\ref{eq:curpol3}), to obtain $\langle j_{5\mu}\rangle ={\kappa\over 8\pi}\epsilon_{\mu\nu\lambda}\partial_\nu\partial_\lambda(\chi_+-\chi_-)$. 
This implies that elementary vortices assigned to different partitions carry 
opposite axial charge  $Q_5=\int d^2\vex{x} \langle j_{50}\rangle=\pm{\kappa/4}$.
 If we had a chiral gauge field probe at our disposal, we could in principle detect the axial charge 
associated with a vortex just as we can detect its electric charge. In the 
absence of such a probe the axial charge still can affect the physics, e.g., by 
influencing the exchange statistics.

The Lagrangian (\ref{eq:l1}) posesses an emergent global symmetry under the 
transformation 
$\psi\to e^{i\chi_0\gamma_5}\psi$ and $\chi\to\chi-2\chi_0$. This guarantees
conservation of the axial charge in all low-energy processes. When a vortex is
created, e.g., in a pair-creation process, it is endowed 
by a particular value of the electric charge $(Q=\pm{1\over 2})$ and the axial 
charge $(Q_5=\pm{1\over 2})$. These values then uniquely characterize the 
vortex; in particular they determine its exchange statistics.
In order to better appreciate this key point 
it is useful to recall that $\psi_\pm$ represent the field operators associated 
with the two different Dirac nodes~\cite{HouChaMud07a,SerWeeFra07a}. 
Physical meaning of the axial charge then becomes clear from the expressions
$j_0=\psi^\dagger_+\psi_++\psi^\dagger_-\psi_-$ and
$j_{50}=\psi^\dagger_+\psi_+-\psi^\dagger_-\psi_-$. 
Since the nodes interchange under $\cal{T}$ and $\cal{P}$ the axial charge is 
{\em odd} under these operations. The spatial components of 
$ j_5$ are closely related to the `valley currents' recently studied
in the context of graphene~\cite{rycerz1}.

The above result (\ref{theta}) could also be understood using the following 
physical picture. Starting with two vortices in the \mbox{$+$-partition} we 
could choose to transfer one, say vortex 2,  to the \mbox{$-$-partition}. The 
symmetric field, $b$, does not change but the anti-symmetric field,
$a$, shifts by $-u_2$. This shift could be absorbed in $A$, which will then 
attach a $2\pi$ flux to vortex 2. As a result, we pick up an extra Aharonov-Bohm
phase $\pi$ by taking the flux $2\pi$ of vortex 2 around the charge 
$\frac12$ of vortex 1. The statistics remains unchanged.

\emph{Doubled CS theory} -- It is possible to rewrite the  Lagrangian
(\ref{eq:lcs2}) in a more familiar form by introducing two auxiliary gauge fields
$\mathcal{A}_\pm$ mediating the statistical interaction between vortices,
\begin{eqnarray}
\mathcal{L}_{\mathrm{CS}}&=&\sum_{\sigma=\pm}
\left(-\frac{1}{\pi\kappa}\sigma\mathcal{A}_\sigma\cdot\partial\times \mathcal{A}_\sigma+j_\sigma\cdot \mathcal{A}_\sigma\right)
\label{eq:lcs3}
\end{eqnarray}
where $j_\pm={1\over 2\pi}(\partial\times\partial\chi_\pm)$ are currents
associated with vortices in two partitions. Gaussian integration over
$\mathcal{A}_\pm$ leads back to Eq.~(\ref{eq:lcs2}).  We recognize the above
Lagrangian as the level-2 doubled CS theory of Ref.~\onlinecite{freedman1},
constructed there based on very general considerations.

\emph{Concluding remarks} -- The statistical angle $\pi/2$ of a vacant vortex means 
that fusing two vacant vortices must result in a boson (with statistical angle 
$2\pi$). At first sight this seems counterintuitive, since the resulting charge 
is $\frac12+\frac12=+1$ which could be expected to be a regular hole, and thus 
a fermion. But there is a subtlety here. Fusing two vortices gives a 
double vortex which supports \emph{two exact zero modes}~\cite{HouChaMud07a,SerWeeFra07a,JacRos81a}. Now, suppose we remove the hole by adding an electron, so 
we have a neutral vortex. Since the electron can go to either of the
two modes there is a pseudospin-${1\over2}$ degree of freedom attached to the neutral 
vortex. By the spin-statistics theorem this indeed should be a
\emph{fermion}, consistent with the notion that the original double vortex
is a charge-1 boson.

It is also interesting to contemplate exactly how our model evades the argument 
of Ref.~\onlinecite{kivelson1}. The crucial point is that since the axial charge 
is odd under time reversal the many-body ground state $\Psi$ of the system 
with two vortices must be at least two-fold degenerate, with 
$\Psi^*=\mathcal{T}\Psi$ 
and $\Psi$ mutually orthogonal. However, the absence of anyons in a 
$\mathcal{T}$-invariant system only follows from Ref.~\onlinecite{kivelson1}
when the state is {\em non-degenerate}. It is straightforward to generalize the
argument to the doubly degenerate situation~\cite{ser1}. One finds that because
of the additional structure introduced by the degeneracy, 
$\theta=\pm\pi/2$ semions are allowed in addition to fermions and bosons,
in agreement with the results of the effective theory.

While this work was in the final stages a preprint by Chamon 
{\it et al.}~\cite{ChaHouJac07a} appeared, in which the O(3) topological current
is derived consistent with our Eq.~(\ref{eq:O3cur}). Their conclusions
regarding the exchange statistics, however, appear to disagree with ours.

\emph{Acknowledgment} -- The authors have benefited from discussions 
with C. Chamon, M.P.A. Fisher, R. Jackiw, E.-A. Kim, S.A. Kivelson, 
M. Lawler, S.H. Simon, G. Semenoff and K. Shtengel.
This research has been supported by NSERC, CIAR
NSF Grant No. PHY05-51164 (KITP) and the Killam Foundation. M.F. acnowledges
hospitality of the Aspen Center for Physics.

\end{document}